# Reliable operation of Cr$_2$O$_3$:Mg/ $\beta$-Ga$_2$O$_3$ p-n heterojunction diodes at 600°C

William A. Callahan[1,2,*], Kingsley Egbo[1], Cheng-Wei Lee[3], David Ginley[1], Ryan O'Hayre[3], Andriy Zakutayev[1,†]

[1] Materials Science Center, National Renewable Energy Laboratory, Golden, Colorado 80401, USA

[2] Advanced Energy Systems Graduate Program, Colorado School of Mines, Golden, Colorado, 80401, USA

[3] Department of Metallurgical and Materials Engineering, Colorado School of Mines, Golden, Colorado, 80401, USA

* Will.Callahan@nrel.gov

† Andriy.Zakutayev@nrel.gov

Abstract

$\beta$-Ga$_2$O$_3$-based semiconductor heterojunctions have recently demonstrated improved performance at high voltages and elevated temperatures and are thus promising for applications in power electronic devices and harsh-environment sensors. However, the long-term reliability of these ultra-wide band gap (UWBG) semiconductor devices remains barely addressed and may be strongly influenced by chemical reactions at the p-n heterojunction interface. Here, we experimentally demonstrate operation and evaluate the reliability of Cr$_2$O$_3$:Mg/ β-Ga$_2$O$_3$ p-n heterojunction diodes at during extended operation at 600°C, as well as after 30 repeated cycles between 25-550°C. The calculated pO$_2$-temperature phase stability diagram of the Ga-Cr-O material system predicts that Ga$_2$O$_3$ and Cr$_2$O$_3$ should remain thermodynamically stable in contact with each other over a wide range of oxygen pressures and operating temperatures. The fabricated Cr$_2$O$_3$:Mg / $\beta$-Ga$_2$O$_3$ p-n heterojunction diodes show room-temperature on/off ratios >10$^4$ at $\pm$5V and a breakdown voltage (V$_{Br}$) of -390V. The leakage current increases with increasing temperature up to 600°C, which is attributed to Poole-Frenkel emission with a trap barrier height of 0.19 eV. Over the course of a 140-hour thermal soak at 600°C, both the device turn-on voltage and on-state resistance increase from 1.08V and 5.34 mΩ-cm$^2$ to 1.59V and 7.1 mΩ-cm$^2$ respectively. This increase is attributed to the accumulation of Mg and MgO at the Cr$_2$O$_3$/Ga$_2$O$_3$ interface as observed from TOF-SIMS analysis. These findings inform future design strategies of UWBG semiconductor devices for harsh environment operation and underscore the need for further reliability assessments for $\beta$-Ga$_2$O$_3$ based devices.

Beta gallium oxide (β-Ga₂O₃) is a prime candidate for next-generation sensing and power electronic devices, especially for high-temperature applications. Due to its large bandgap and high theoretical breakdown field, β-Ga₂O₃ shows promise for high-voltage operation and high-frequency switching. Additionally, β-Ga₂O₃ can be doped n-type to relatively high levels, most commonly with Si and Sn. Single-crystal β-Ga₂O₃ can be grown via multiple bulk methods, including Czochralski (CZ) and edge-defined film-fed growth (EFG). Techno-economic analysis and projective cost modeling suggests that further development of Ga₂O₃-based technologies will improve their commercial viability in competition with current SiC and GaN technologies[1–6].

Over the last few years significant progress has been made in expanding the capabilities of β-Ga₂O₃ devices towards high-performance operation, with several groups making efforts to reach the predicted theoretical high-power limits via maximization of breakdown voltage and simultaneous minimization of on-state resistance. This has been achieved through a shift from relatively simple Schottky/heterojunction device structures towards more complex stacks that incorporate field plates, guard rings, and dielectric layers[7]. However, there are few operational demonstrations of these devices at high temperature, specifically above 300-400°C, and even fewer that study performance stability[8]. Previously, we found that a combination of repeated cycling between low and high temperatures ("cycling"), as well as long-term exposure to very high temperatures ("soaking") provides insight into both mechanical and thermodynamic stressors that can degrade device electrical performance[9,10].

A variety of p-type materials have been explored for use in β-Ga₂O₃ device fabrication, including SiC, GaN, NiO, diamond, SnO, and Cu₂O[11–16]. Nickel oxide (NiO) is the most widely explored due to its wide bandgap and favorable band alignment with β-Ga₂O₃, achieving turn-on voltages in the range of 0.9-1.8V and reliably low leakage currents during room-temperature operation. Furthermore, very high breakdown voltages and excellent corresponding Baliga figures of merit have been reported for NiO/ β-Ga₂O₃ heterojunctions[17–27]. Of the candidate p-type metal oxides for use in β-Ga₂O₃ device fabrication, several (e.g., Cu₂O, NiO) are predicted to form ternary compounds when in contact with Ga₂O₃ based on computational thermodynamics[28]. Here we report β-Ga₂O₃ heterojunction diodes using the wide bandgap *p*-type chromium oxide (Cr₂O₃), for which there is virtually no literature. However, recent XPS work on Ga₂O₃/Cr₂O₃ suggest favorable band alignment ($\Delta E_C =$ 1.68 eV; $\Delta E_V =$ 3.38 eV) for heterojunction device fabrication[29,30].

In this work, epitaxial Cr₂O₃:Mg/β-Ga₂O₃ vertical pn-diodes are fabricated, and their electrical performance and thermal reliability are evaluated. We demonstrate the first example of high-performance Cr₂O₃:Mg/β-Ga₂O₃ heterojunction with a room-temperature rectification ratio

>$10^4$ at $\pm 5$V, and breakdown voltage of -390V. We then subject several Cr$_2$O$_3$:Mg/$\beta$-Ga$_2$O$_3$ vertical pn-diodes to repeated thermal cycling between 25-550°C, as well as separately to a 140-hour thermal soak at 600°C. We find that the devices show exceptionally stable reverse bias leakage current during prolonged high-temperature testing, but experience forward-bias changes in the form of increased on-state resistance, turn-on voltage, and forward current drop. The reverse bias leakage current can be described by Poole-Frenkel emission with a trap barrier height of 0.19 eV. We attribute the observed evolution in the forward bias electrical properties to the accumulation of Mg and MgO at the Cr$_2$O$_3$/Ga$_2$O$_3$ interface after prolonged operation at 600°C due to the temperature induced diffusion of the Mg dopants in Cr$_2$O$_3$. This study provides providing new insights and design considerations for oxide-based UWBG semiconductors for p-n heterojunctions applicable in harsh operating environments.

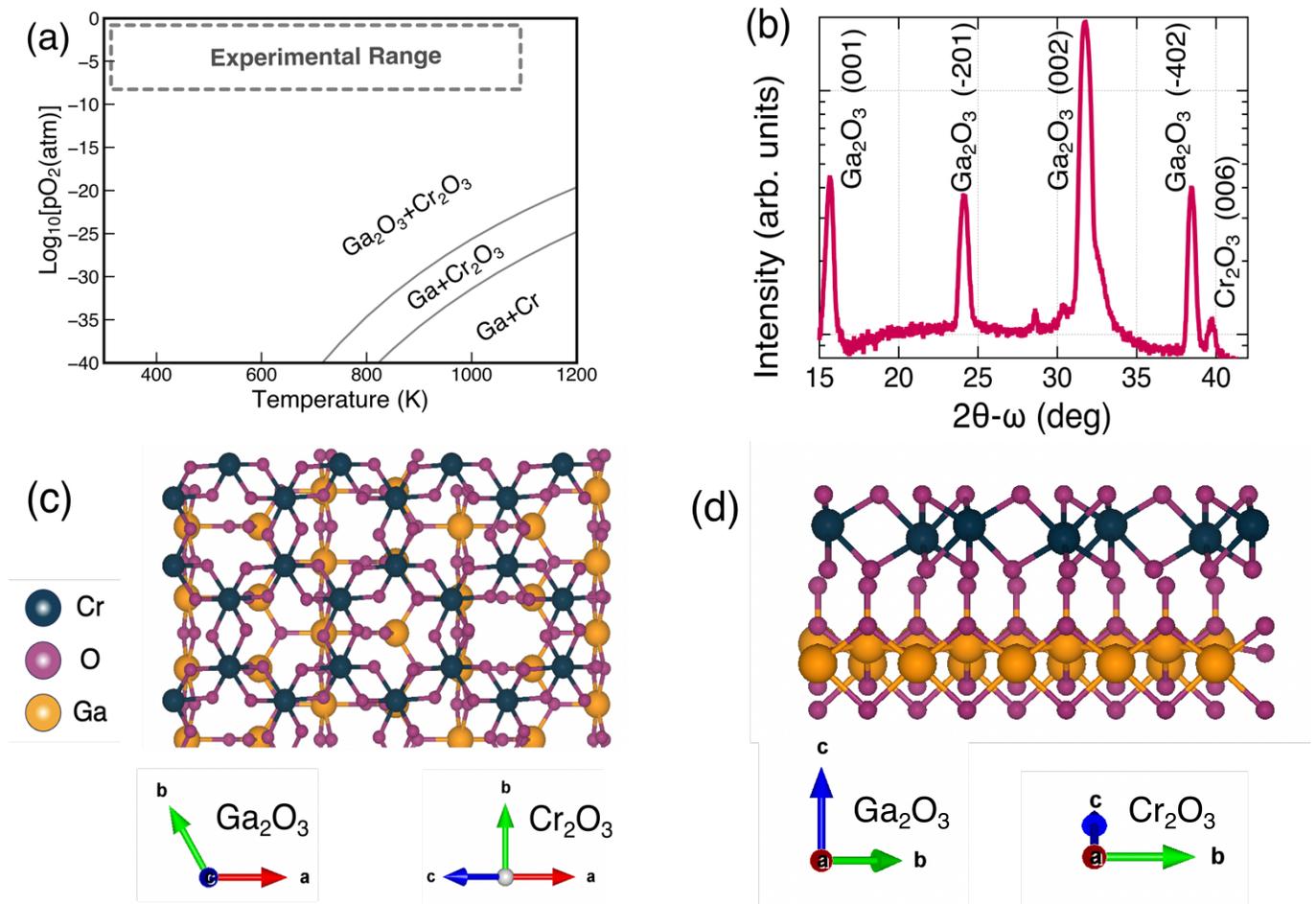

Fig 1: (a) Calculation showing pO$_2$-temperature phase stability diagram for the Ga-Cr-O system. The "Experimental Range" box highlights the expected operating conditions for devices made from these materials, including the conditions used in this report. (b) $2\theta$-$\omega$ scan of a 120 nm Cr$_2$O$_3$:Mg layer grown on (001) HVPE Ga$_2$O$_3$ substrate on a log scale. The region 36-42 deg shows both the (-201) sublattice of the HVPE layer, as well as the preferential growth of (006) Cr$_2$O$_3$ on that sublattice. (c) Top-down and (d) profile views of the (001) Cr$_2$O$_3$ – (001) Ga$_2$O$_3$ interface.

Figure 1(a) shows the calculated oxygen partial pressure *vs.* temperature ($pO_2$-$T$) phase diagrams for the Ga-Cr-O system. The "Experimental Range" box includes the temperature and $pO_2$ conditions used in this report, which are limited by instrumentation. $Cr_2O_3$ and $Ga_2O_3$ are predicted to coexist for effectively all desirable temperatures and oxygen partial pressures. This $pO_2$-$T$ phase diagram for the Ga-Cr-O system was calculated using density functional theory (DFT) datasets from NREL MatDB[31,32] along with correction schemes[33,34] to achieve chemical accuracy (~ 50 meV/atom). Specifically, we estimate the Gibbs formation energy of a compound, $\Delta G_f(T)$, in units of eV/atom by,

$$\Delta G_f(T) = \Delta H_f(298.15K) + G_{SISSO}^{\delta}(T) - \sum_{i}^{N} \alpha_i\, G_{i,exp}(T) \qquad (1)$$

where $\Delta H_f(298.15K)$ is the standard formation enthalpy of the compound and is calculated using DFT total energies from the NREL MatDB and fitted elemental-phase reference energies. Here, $\alpha_i$ is the stoichiometric weight for element $i$ in the compound using experimental values from FactSage[35] for absolute Gibbs free energy of element $i$. $G_{SISSO}^{\delta}(T)$ (eV/atom) captures temperature dependency of the compound due to phonons[34]. Lastly, we collect $\Delta G_f(T)$ for all competing phases and use convex hull analysis to calculate the grand potential phase diagram. The ideal gas law, e.g., $\Delta \mu_O = \frac{1}{2} k_B T \ln(pO_2)$, is used to connect oxygen chemical potential ($\mu_O$) to oxygen partial pressure ($pO_2$) in units of atm, where $\Delta \mu_O$ is the deviation from the standard oxygen chemical potential ($\mu_O^\circ$) with $\mu_O = \mu_O^\circ + \Delta \mu_O$ [36]. A full description of these calculations is available online[37].

$Cr_2O_3$:Mg was deposited via pulsed laser deposition (PLD). During PLD growth, a ceramic target of 8at%Mg-doped $Cr_2O_3$ was ablated using a pulsed KrF excimer laser at a frequency of 10Hz and energy of 300mJ. Growth was performed at a substrate temperature of 600°C and $O_2$ partial pressure of $3 \times 10^{-4}$ Torr. To fabricate Ohmic contacts (Fig S1), $Cr_2O_3$:Mg was grown on a c-sapphire substrate, followed by a bi-metal deposition of 5 nm Ti/ 100 nm Au. E-beam evaporation of metals was performed with a Temescal FC2000 evaporation system under high vacuum conditions ($\leq 3 \times 10^{-6}$ Torr) without venting between layers. Room temperature Hall measurements were performed with a Lakeshore FastHall system. The $Cr_2O_3$:Mg films were determined to be p-type with a carrier concentration of $2.4 \times 10^{17}$ cm$^{-3}$; a resistivity of 31.5 Ω cm, which is in good agreement with reports on Mg-doped $Cr_2O_3$[38,39]; and a mobility of 0.82 cm$^2$ V$^{-1}$s$^{-1}$, which can be significantly improved with choice of dopant[40]. High-temperature Hall measurements were also attempted using a custom instrument[41], but the results were inconclusive. To electrically test the Ohmic contacts to $Cr_2O_3$:Mg, two-probe measurements (source I, measure V) were performed via a Keithley 236 SMU in a custom McAllister probe station in ambient air. IV curves were collected every 15 minutes. To initiate the long-term thermal soak, the probe station was ramped to 600°C at a rate of 100°C per hour. Over the course of the one week testing period at 600°C, the resistance remains effectively constant, and the IV curves remain linear. Electrical measurements reveal that Cr$_2$O$_3$:Mg is a highly resistive semiconductor, especially at lower temperatures. Full results and figures for the Ohmic contact testing can be found in the Supplementary Material Fig. S1.

Figure 1(b) gives a $2\theta$-$\omega$ scan of a 120 nm Cr$_2$O$_3$:Mg layer grown on (001) Si-doped HVPE Ga$_2$O$_3$ substrate, performed using a Bruker 8 Discover diffractometer. In addition to the (001) Ga$_2$O$_3$ planes, there are also peaks that indicate the presence of a (-201) Ga$_2$O$_3$ sublattice within the HVPE layer. In the region from 36 to 42 degrees, the (006) plane of Cr$_2$O$_3$ can be observed and its proximity to the (-402) plane of Ga$_2$O$_3$ suggests preferential growth on this sublattice. This relationship is further supported by Figure S4, which shows a similar $2\theta$-$\omega$ scan of Cr$_2$O$_3$:Mg layer preferentially growing in the (001) orientation on a reference (-201) $\beta$-Ga$_2$O$_3$ substrate. Figures 1(c) and 1(d) give a top-down and profile visualization of the stacked (001) planes for both materials in the Cr$_2$O$_3$-Ga$_2$O$_3$ relationship, respectively.

Cr$_2$O$_3$:Mg/$\beta$-Ga$_2$O$_3$ p-n vertical heterojunction devices were fabricated using 1 $\mu m$ thick $\beta$-Ga$_2$O$_3$ drift layers (n $\cong 4 \times 10^{16}$cm$^{-3}$) grown via HVPE on (001) Ga$_2$O$_3$:Sn from Novel Crystal Technologies. Photoresist was removed from the as-delivered substrates via an organic wash followed by a sulfuric acid/peroxide rinse. Next, large-area back Ohmic contacts (5 nm Ti / 100 nm Au) and the p-Cr$_2$O$_3$:Mg layer were deposited via e-beam and PLD, respectively, via the methods previously described. Two different geometries were used for the top Ohmic contacts to Cr$_2$O$_3$:Mg. For the thermal cycling devices, shadow masks were used to define an array of 1 mm diameter pads that were sufficiently large to ensure reliable probe contact despite repeated thermal expansion and contraction. To fabricate devices for the long-term thermal soak, contact aligner lithography was used to make circular pads of increasing size, from 50-300 $\mu m$ in diameter. In both cases, the 5 nm Ti/100 nm Au contacts were deposited as before. Following contact deposition, both Ohmic and heterojunction samples were annealed a single time via rapid thermal processing (ULVAC-RIKO MILA-3000 Rapid Thermal Processing Unit) at 550°C for 90s in flowing N$_2$.

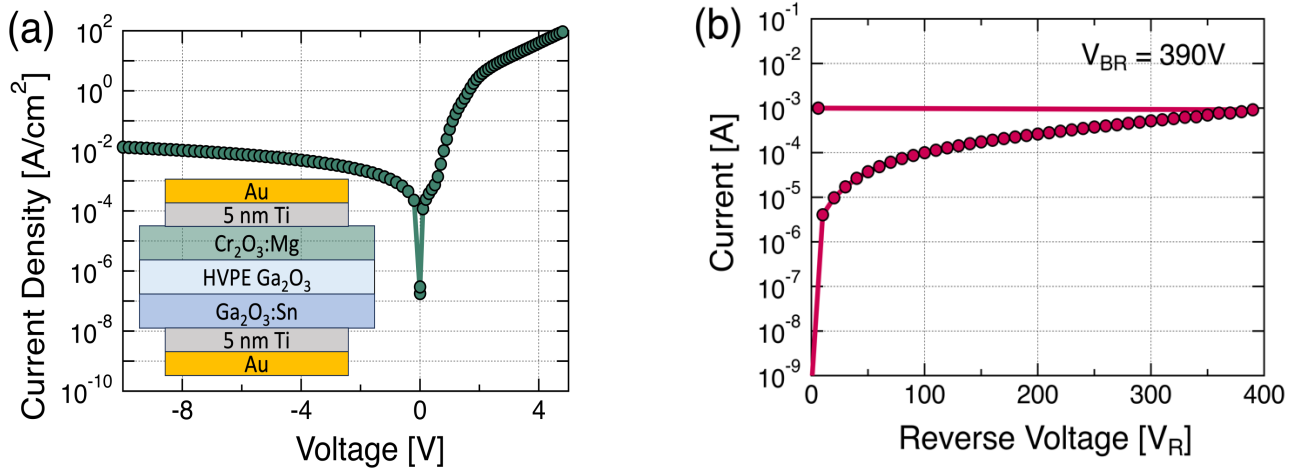

Figure 2: (a) Room temperature JV characteristics for 300 $\mu m$ Cr$_2$O$_3$:Mg/$\beta$-Ga$_2$O$_3$ p-n heterojunction diode, shown in forward bias on a linear scale. Insert shows the device architecture of the p-n heterojunctions used in this report. (b) Room-temperature reverse bias breakdown performance.

Figure 2 gives the room-temperature electrical characteristics of a 300 $\mu m$ diameter Cr$_2$O$_3$:Mg / $\beta$-Ga$_2$O$_3$ p-n diode under 50 sccm flowing Ar. Fig 2(a) shows the entire JV characteristic on a log scale, measured from +5 to -10V. A rectification ratio greater than $10^4$ is achieved at $\pm5$V. The diode reaches a turn-on current density of 1 A/cm² at approximately 1.65V and fitting of the "On" portion of the curve on a linear scale (Fig S2) yields an $R_{on}$ of 12.96 m$\Omega$ cm². The inset shows the device p-n heterojunction architecture for devices studied in this report. Details about doping levels and layer thicknesses can be found in Fig. S2. Figure 2(b) gives a reverse bias breakdown voltage of 0.39 kV. Room-temperature breakdown voltage was measured using a Keysight B1505A Power Device Analyzer with a reverse current limit of 1 mA. This device is among the first demonstrations of a Cr$_2$O$_3$:Mg/ $\beta$-Ga$_2$O$_3$ p-n diode; significant further improvements can be expected based on architectural enhancements to control the electric field and optimize the growth conditions.

Both thermal cycling and soaking of the heterojunction diodes were performed with a Keithley 2635A SMU in an Instec HCP621G-PMH temperature-controlled probe station. Thermal cycling consists of repeated ramping from 25-550°C and back down in steps of 75°C. The instrument was allowed to equilibrate for 15-20 minutes at each temperature before electrical measurements were collected. Thermal soaking was performed in a similar fashion; the instrument was manually ramped from 25°C to 600°C in steps of 75-100°C. The instrument was allowed to equilibrate for 20-30 minutes at each temperature, after which electrical measurements were collected. Upon reaching 600°C, electrical measurements were performed every 30 minutes for the full duration of the soaking experiment. For thermal cycling, IV curves were collected from $0 \rightarrow +5$V, followed by a reverse bias sweep from $0 \rightarrow -20$V.

Compliance was set to 0.1A for all measurements. For thermal soaking electrical measurements, IV curves were collected from 0 → +5V with a 0.2A current limit, followed by a reverse bias sweep from 0 → −10V with a 0.1A current limit.

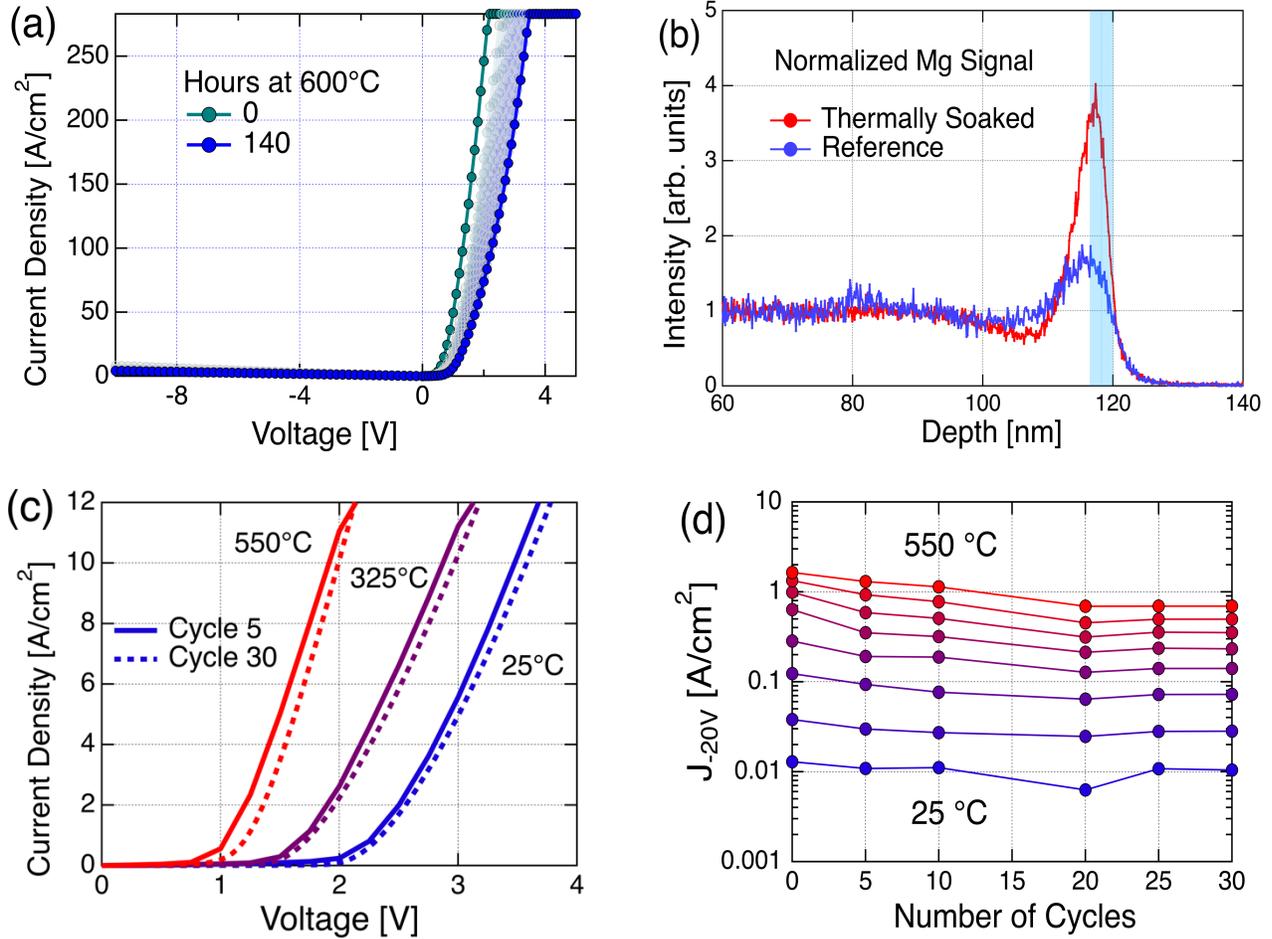

Figure 3: (a) Linear JV curves of a 300 μm diameter Cr$_2$O$_3$:Mg/ β-Ga$_2$O$_3$ p-n diode operating at 600°C for 140 hours, with traces plotted every 10 hours. The current limits were set to 200 mA in forward bias and 100 mA in reverse bias to prevent damage during operation. (b) normalized Mg signal from TOF-SIMS for the thermally soaked diode and an as-deposited reference device. The Ga$_2$O$_3$/Cr$_2$O$_3$ interface is indicated by the blue rectangle. The signals have been normalized relative to their baseline signal taken from between 20-80 nm in the Cr$_2$O$_3$ layers. (c) Forward bias JV curves of a 1 mm diameter Cr$_2$O$_3$:Mg/β-Ga$_2$O$_3$ pn diode for three select temperatures before and after repeated thermal cycling. (d) Evolution of temperature dependent leakage current with repeated cycling, measured at -20V.

Figure 3 shows the high-temperature reliability of several Cr$_2$O$_3$:Mg/ β-Ga$_2$O$_3$ pn diodes under both thermal soaking and cycling conditions. Fig 3(a) shows results of a >100-hour thermal soak at 600˚C for a 300 μm Cr$_2$O$_3$:Mg/ β-Ga$_2$O$_3$ pn diode under 50 sccm flowing Ar. JV curves were measured every 30 minutes, with traces plotted every 10 hours to demonstrate the time evolution. A line was fit to the "On" portion of Fig. 3(a) from 100 to 250 A/cm² and extrapolated to determine the turn-on voltage and specific on-state resistance. After a break-in period of 10 hours, $V_{on}$ steadily increases from 1.08V to 1.59V after 140 hours, at a rate of approximately 4 mV per hour. Correspondingly, the value of $R_{On}$ also increases from 5.34 mΩ cm² to 7.1 mΩ cm² over the same period but appears to grow with the square root of time. The evolution of reverse bias leakage currents was also examined and found to stabilize at all voltages after a break-in period of ~30 hours. Plots of $R_{On}$, $V_{On}$, and $J_{Rev}$ vs. time can be found in Fig. S2(c-e). Fig S2(f) shows the forward-bias JV curves for the same 300 μm $Cr_2O_3$:Mg/ β-$Ga_2O_3$ pad taken before ("Pre", solid lines) and after ("Post", dashed lines) the 140-hour soak at 600°C, measured out to +5V. Significant changes in performance can be observed, especially at lower temperatures. Room-temperature $V_{On}$ and $R_{On}$ increase from 1.65V and 12.96 mΩ cm² to 3.84V and 312 mΩ cm², respectively.

To understand the cause of the increase in forward drop, both the thermally soaked ("test") sample and an as-deposited ("reference") sample were studied using time-of-flight secondary ion mass spectrometry (TOF-SIMS) analysis. Results are shown in Fig 3(b) and Fig S5. The test sample reveals a significant amount of Mg dopant diffusing to the $Cr_2O_3$/$Ga_2O_3$ interface after thermal soaking at 600°C. Beyond this, there appears to be no interdiffusion of any other species between the p-type and n-type layers – the interface remains abrupt, which supports our theoretical calculations. The normalized Mg and MgO signals at the interface from the test and reference sample are also compared. Both signals at the $Cr_2O_3$/$Ga_2O_3$ interface in the test sample are significantly higher than in the reference sample. While there is evidence of a small amount of Mg accumulation in the reference sample, very little of it appears to have oxidized. This diffusion can be attributed to the short-term high-temperature growth conditions experienced during PLD. MgO is a well-known wide bandgap material that is frequently used for its dielectric strength[46]. Given the band alignment of $Ga_2O_3$ with MgO and $Cr_2O_3$[29,47], this observed reaction amounts to inserting a thin dielectric layer between the p-type and n-type layers. Assuming the thickness of this MgO layer is increasing with time at temperature, this potentially explains the gradual increase in both forward bias resistance and turn-on voltage observed in the electrical measurements. The reverse bias leakage current stabilizes after 20-40 hours, which suggests that the mechanism for current transport has a relatively low energy barrier and isn't strongly dependent on the thickness of the Mg/MgO layer. While the final recorded turn-on voltage at 600°C approaches the band offset of $Cr_2O_3$ and $Ga_2O_3$ measured with XPS[29,30], it's more likely that this attributable to the MgO layer. Temperature-dependent XPS measurements are needed to better understand this relationship.

A separate device utilizing 1 mm diameter pads was subject to repeated thermal cycling between 25-550°C, with current compliance set to 100mA. Forward bias curves were measured out to 5V; reverse bias was measured to -20V. Figure 3(c) shows the forward-bias J-V-T curves for cycle 5 and cycle 30 for three different temperatures, showing slight increases in $V_{on}$ with negligible changes in $R_{on}$. Figure 3(d) shows the evolution of leakage current with increasing cycle, as measured at -20V. With a few exceptions at lower temperature, the leakage current appears to stabilize after ~20 cycles.

Overall, Figure 3 demonstrates the different effects of thermal cycling and thermal soaking on the performance of electronic devices. While both impart a significant thermal load, the 140-hour soak at 600°C appears to have greater impact on device recovery, especially at lower temperatures. In contrast, the changes in electrical behavior between cycles 5 and 30 in the temperature cycling test are significantly less severe. In both experiments, the reverse bias leakage current stabilizes after a somewhat significant break-in period (~30 hours or 20 cycles, respectively); the forward bias current appears to be more affected in both cases. Another consideration is the size of the pad being measured. Given the poor thermal conductivity of $\beta$-Ga$_2$O$_3$, it is possible that device performance degradation can be attributed to substantial Joule heating. Since there is a greater than 10-fold increase in the area of the thermally cycled devices when compared to those that were soaked, the generated heat can be more evenly distributed across the contact area.

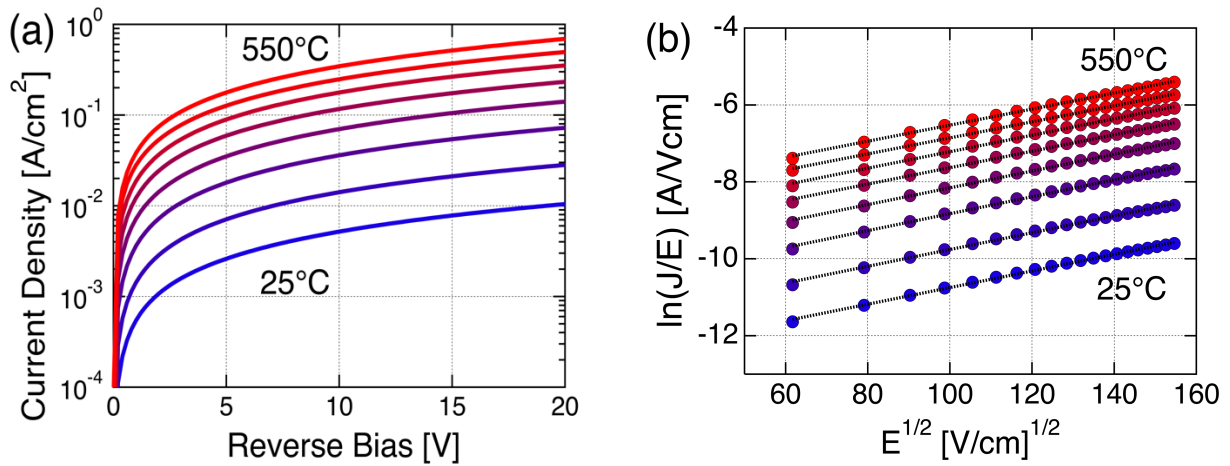

Figure 4: (a) Reverse bias JV curves from cycle 30, showing dependences on both temperature and voltage. (b) Poole-Frenkel plots of leakage current for cycle 30, with linear fits to each temperature trace shown in dotted black lines.

To understand the corresponding leakage current mechanism, we examined reverse bias J-V-T measurements from cycle 30 (Fig 4(a)). The results show that reverse bias current has a strong dependence on both temperature and voltage. To analyze these results, we use the Poole-Frenkel emission (PFE) model, which describes conduction via thermal excitation of electrons from trap states and is typically present in insulating layers. PFE can expressed by the following equation:

$$J_0 \propto E \exp\left(-\frac{q(\phi_F - \sqrt{qE/\pi\varepsilon_s\varepsilon_0})}{k_B T}\right) \quad (2)$$

Here, $E$ is the electric field across the depletion region, $\phi_F$ is the barrier height of the trap state, $\varepsilon_s$ is the relative dielectric constant, and $k_B$ is Boltzmann's constant. To extract the barrier height of these trap states, graphs of $\ln\left(\frac{J}{E}\right)$ vs. $\sqrt{E}$ are first plotted for each temperature, showing excellent linear fits (Fig 4(b)). Subsequently, the intercepts of these linear fits are plotted against $q/k_B T$ (Fig S3), yielding trap barrier heights of 0.191 eV[42–44]. This relatively low barrier height is smaller than what has been reported for NiO/Ga$_2$O$_3$[45], which results in the leakage current of the Cr$_2$O$_3$:Mg/$\beta$-Ga$_2$O$_3$ diodes that is several orders of magnitude higher than NiO/Ga$_2$O$_3$ heterojunctions.

The J-V-T results and analysis presented in Figure 4 indicate that the cycling and soaking reliability trends presented in Figure 3 may be related to non-ideal interfaces in the overall Au/Ti/Cr$_2$O$_3$:Mg/$\beta$-Ga$_2$O$_3$/Ti/Au multi-layer device stack (Figure 1a, inset). The contribution of Au/Ti/Cr$_2$O$_3$:Mg or $\beta$-Ga$_2$O$_3$/Ti/Au Ohmic contacts to this device degradation can be ruled out. This is because the independent Au/Ti/Cr$_2$O$_3$:Mg/Al$_2$O$_3$ Ohmic contacts stability study (see Fig. S1) showed that they remain stable for a longer time than the device as a whole. Our earlier published results on $\beta$-Ga$_2$O$_3$ showed stable Ohmic behavior after >500 hours of continuous operation at 600°C[9]. Another potential interface where degradation can occur is Cr$_2$O$_3$:Mg/$\beta$-Ga$_2$O$_3$. As shown in Figure 1(b), there is a quasi-ordered relationship between the Cr$_2$O$_3$ layer and the (-201) character of the (001)-oriented HVPE Ga$_2$O$_3$ substrate. The lack of direct epitaxy can result in a lower interface quality, which can lead to a high density of interfacial trap states and thereby facilitate reverse-bias conduction. Furthermore, this might be exacerbated by the accumulation of Mg and MgO at this interface. However, this is unlikely because similar lack of ideal epitaxy exists at the NiO/$\beta$-Ga$_2$O$_3$ interface which shows higher barrier extracted from the J-V-T measurement using PFE model[45]. It is worth noting that based on the minimal changes to forward bias conduction and the shorter cumulative exposure time to high temperatures, we do not anticipate as much Mg/MgO migration in the thermally cycled samples as in the thermally soaked samples.

In this work, we demonstrate reliable operation of Cr$_2$O$_3$:Mg/ β-Ga$_2$O$_3$ p-n heterojunction diodes at 600°C. This experimental work is motivated by theoretical calculations that predict thermodynamically stable phase coexistence of n-Ga$_2$O$_3$ and p-Cr$_2$O$_3$ for an extremely wide range of temperatures and pO$_2$. We fabricate some of the first reported Cr$_2$O$_3$:Mg/$\beta$-Ga$_2$O$_3$ vertical p-n heterojunction diodes with $V_{on}$, $R_{on}$, and $V_{br}$ of 1.65V, 12.96 mΩ cm$^2$, and -390V, respectively at room-temperature. We subject several diodes to repeated thermal cycling between 25-550°C and long-term continuous operation at 600°C for 140 hours. Analysis of the reverse bias current shows that the leakage mechanism is very well described by Poole-Frenkel emission. We find that prolonged thermal soaking has more deleterious effects on electrical performance than repeated cycling, leading to increased turn-on voltage and on-state resistance. Since we observed stability of the ultrathin Ti/Au Ohmic contact to $Cr_2O_3$:Mg/c-$Al_2O_3$ as well as $Ga_2O_3$/Ti/Au contacts under the same high-temperature operating conditions, we attribute this performance evolution to the migration of Mg to the $Cr_2O_3$/$Ga_2O_3$ interface, where SIMS reveals the presence of a thin MgO layer. Future work should focus on understanding these degradation phenomena and their temperature dependence, as well as comparisons to other material systems under similar conditions.

Supplementary Material Description

The supplementary material contains:

- S1) Device architecture for Ohmic contact devices. Temperature-dependent performance of Ohmic contact.
- S2) Room-temperature and high-temperature JV curves of $Cr_2O_3$:Mg/$\beta$-$Ga_2O_3$ p-n devices, as well as extracted parameters from a 140-hour soak at 600°C.
- S3) Poole-Frenkel barrier height extraction.
- S4) XRD of (001) $Cr_2O_3$:Mg / (-201) $\beta$-$Ga_2O_3$.
- S5) TOF-SIMS results for thermally soaked and reference $Cr_2O_3$:Mg / $\beta$-$Ga_2O_3$ samples.

Author Declarations

The authors have no conflicts to declare.

Data Availability

The data supporting the findings of this study are available within the paper and its Supplementary Material. Any additional data connected to the study are available from the corresponding author upon reasonable request.

Acknowledgements

This work was authored by the National Renewable Energy Laboratory (NREL), operated by Alliance for Sustainable Energy, LLC, for the U.S. Department of Energy (DOE) under Contract No. DE-AC36-08GO28308. Funding is provided by the Office of Energy Efficiency and Renewable Energy (EERE) Advanced Manufacturing Office. The views expressed in the article do not necessarily represent the views of the DOE or the U.S. Government. This material makes use of the TOF-SIMS system at the Colorado School of Mines, which was supported by the National Science Foundation under Grant No.1726898. We would especially like to thank V.S. for his contributions to the computational methods used in this report; C.E.P for her assistance in high-temperature transport measurements; and M.A.W for his analysis with the TOF-SIMS system.

Author Contributions

William A. Callahan – Conceptualization (equal), Data Curation (lead), Formal Analysis (lead), Methodology (lead), Software (equal), Validation (lead), Visualization (equal), Writing/Original Draft Preparation (lead).
Kingsley Egbo– Conceptualization (equal), Resources (equal), Writing – review & editing (equal)
Cheng-Wei Lee– Software (equal), Visualization (equal), Writing – review & editing (equal)
Ryan O'Hayre: Funding acquisition (equal), Writing – review & editing (equal), Supervision (equal)
David Ginley: Funding acquisition (equal), Writing – review & editing (equal), Supervision (equal)
Andriy Zakutayev: Project administration (lead), Supervision (equal), Funding acquisition (equal), Writing – review & editing (equal)